\documentclass[aps,pra,twocolumn,superscriptaddress,byrevtex,showpacs,preprintnumbers,amsmath,amssymb,floatfixm]{revtex4}

\usepackage{graphicx,hyperref}
\usepackage{dcolumn}% Align table columns on decimal point
\usepackage{bm}% bold math
\usepackage{color}
\usepackage{wasysym}
\usepackage{amssymb, amsfonts,amsmath,amstext,revsymb}

\begin{document}
\title{Producing translationally cold, ground-state CO molecules}
\author{Janneke H. Blokland}
\author{Jens Riedel}
\author{Stephan Putzke}
\affiliation{Fritz-Haber-Institut der Max-Planck-Gesellschaft,
Faradayweg 4-6, 14195 Berlin, Germany}
\author{Boris G. Sartakov}
\affiliation{General Physics Institute, RAS, Vavilov Street 38,
119991 Moscow, Russia}
\author{Gerrit C. Groenenboom}
\affiliation{Theoretical Chemistry, Institute for Molecules and
Materials, Radboud University Nijmegen, Heyendaalseweg 135, 6525 AJ
Nijmegen, The Netherlands}
\author{Gerard Meijer}
\email[Electronic address: ]{meijer@fhi-berlin.mpg.de}
\affiliation{Fritz-Haber-Institut der Max-Planck-Gesellschaft,
Faradayweg 4-6, 14195 Berlin, Germany}

\date{\today}

\begin{abstract}
Carbon monoxide molecules in their electronic, vibrational, and
rotational ground state are highly attractive for trapping
experiments. The optical or ac electric traps that can be envisioned 
for these molecules will be very shallow, however, with depths in
the sub-milliKelvin range. Here we outline that the required samples 
of translationally cold CO (X$^1\Sigma^+$,~$v''$=0,~$N''$=0) 
molecules can be produced after Stark deceleration of a beam of 
laser-prepared metastable CO (a$^3\Pi_1$) molecules followed by 
optical transfer of the metastable species to the ground state 
\emph{via} perturbed levels in the A$^1\Pi$ state. The optical transfer
scheme is experimentally demonstrated and the radiative lifetimes
and the electric dipole moments of the intermediate levels are 
determined.
\end{abstract}


\maketitle

\section{Introduction}

During the last decades, researchers have learned to gain ever
better control over the internal degrees of freedom and the
translational motion of molecules, and a wide variety of molecules
have been trapped by now in electric, magnetic or optical fields
\cite{Taylor&Francis-2009,Faraday-142-2009}. The lowest
temperatures and the highest densities have been obtained for
molecules that are assembled from their laser-cooled atomic
constituents, \emph{e.g.}, for homo- and hetero-nuclear alkali
dimers \cite{Ni-Science-2008,Deiglmayr-PRL-2008}. Cooling and
trapping of pre-existing molecules often involves highly reactive
species that are produced, for instance, via laser ablation or
photodissociation of a precursor molecule. The production
efficiency of such radical species limits the absolute number and
the number density that can be reached for trapped samples. In a
cooling and trapping experiment, one would ideally like to use a
stable molecule that is available in a bottle at high vapor
pressure. Among all the molecules that have been trapped so far,
only ammonia \cite{Bethlem-Nature-2000}, carbon monoxide
\cite{Gilijamse-JCP-2007} and  molecular hydrogen \cite{Seiler-PRL-2011} fall in this category. However, the
various symmetric top isotopologues of ammonia that have been
trapped have a rather complicated energy-level structure due to
the presence of four atoms with nonzero nuclear spin. Even the
simplest isotopologue, $^{15}$NH$_3$, has its population
distributed over 12 $|$M$_F|$ components when it is trapped in its
electronic, vibrational, and rotational ground state by means of
electric fields. The CO molecules have thus far only been trapped
in their metastable a$^3\Pi$ state. In this open-shell
electronically excited state, electrostatic trapping is possible
by virtue of the electric dipole moment of $+$1.37 D together
with the first-order Stark interaction, but phosphorescence back
to the ground state limits the trapping time to the millisecond
range \cite{Gilijamse-JCP-2007}. Similarly, H$_2$ molecules have thus far only been trapped in short-lived Rydberg states \cite{Seiler-PRL-2011,Seiler-PCCP-2011}

Carbon monoxide in its electronic, vibrational and rotational ground
state (X$^1\Sigma^+$,~$v''$=0,~$N''$=0) is in many ways an ideal
candidate for trapping experiments. The six
stable CO isotopologues combining $^{12}$C or $^{13}$C with
$^{16}$O, $^{17}$O or $^{18}$O are all commercially available. The
most abundant isotopologue $^{12}$C$^{16}$O is a boson with no
nuclear spin. The nondegenerate $N''$=0 level is separated by almost
4 cm$^{-1}$ from the next, $N''$=1, level. Contrary to the case of
many diatomic hydrides \cite{Hoekstra-PRL-2007}, CO is quite immune
to blackbody radiation; even at room temperature, the rate at which
the molecules are optically pumped out of the $N''$=0 level by
blackbody radiation is below 10$^{-3}$ s$^{-1}$.
The $^{12}$C$^{18}$O isotopologue is also
a boson with no nuclear spin, while $^{13}$C$^{16}$O and
$^{13}$C$^{18}$O are fermions with an $F''$=1/2 ground level
($\vec{F} = \vec{N} + \vec{I}$ with $\vec{I}$ the nuclear spin),
which remains a single (doubly degenerate) level in the presence of
an electric field.

Ground-state CO molecules can in principle be trapped in an ac electric 
field trap or in an optical trap, although it is an experimental challenge 
to make a trap with a significant depth. The lowest rotational levels 
of the electronic ground state of CO have a purely quadratic Stark 
shift in electric fields that can be realized in the laboratory. The main 
contribution to the Stark shift is due to the electric dipole moment of 
$-$0.1098 D \cite{Muenter-JMS-1975}, although the isotropic polarizability of the 
ground state \cite{Rijks-JCP-1989} cannot be neglected. Note that we use here the
convention that a positive sign of the dipole moment corresponds to the
situation C$^+$O$^-$, \emph{i.e.}, in the ground-state the situation is C$^-$O$^+$. 
The inset of Fig.~\ref{Energylevels} shows the calculated Stark shifts of 
the three lowest rotational levels of $^{12}$C$^{16}$O for electric fields up 
to 100 kV/cm. It is interesting to note that the acceleration experienced
in electric fields by the CO molecules in the $N''$=0 level is almost 
identical to that experienced by ground-state $^{87}$Rb atoms. 
Therefore, the performance of an ac electric trap for ground-state 
CO molecules will be very similar to that of an ac electric trap for 
ground-state $^{87}$Rb atoms. The latter trap has been demonstrated 
\cite{Schlunk-PRL-2007,Rieger-PRL-2007} and carefully characterized
\cite{Schlunk-PRA-2008} before, from which it is concluded that ac electric
traps with a depth of tens of micro-Kelvin can be made for ground-state
CO molecules. 

In the following sections, we outline and demonstrate a scheme via 
which the required translationally cold samples of ground-state CO 
molecules can be produced. 

\section{Optical pumping scheme}

A high density of CO molecules is obtained in a pulsed beam,
\emph{e.g.} when a mixture of CO in a rare gas carrier is expanded
through a pulsed valve from high pressure into vacuum. In such a
beam, a large fraction of all the CO molecules will be in the
$v''$=0 and $N''$=0 level of the electronic ground state. Even
though the vapor pressure of CO allows the pulsed valve to be
cooled to liquid nitrogen temperatures, the speed of the molecules
in the beam will be several hundred m/s. Deceleration of the CO
molecules can be performed very efficiently when the molecules are
first transferred to the metastable a$^3\Pi$ state. Pulsed laser
excitation around 206 nm is used in the expansion region of the
molecular beam to saturate the spin-forbidden transition to the
metastable state. Molecules in the $J$=1 level of the a$^3\Pi_1$,
$v$=0 state can be decelerated to arbitrarily low velocities; this
has actually been used in the first demonstrations of various types of
Stark decelerators thus far
\cite{Bethlem-PRL-1999,Bethlem-PRL-2002,Meek-PRL-2008,Osterwalder-PRA-2010}.

\begin{figure}[h]
 \centering
     \includegraphics[width=7.5cm]{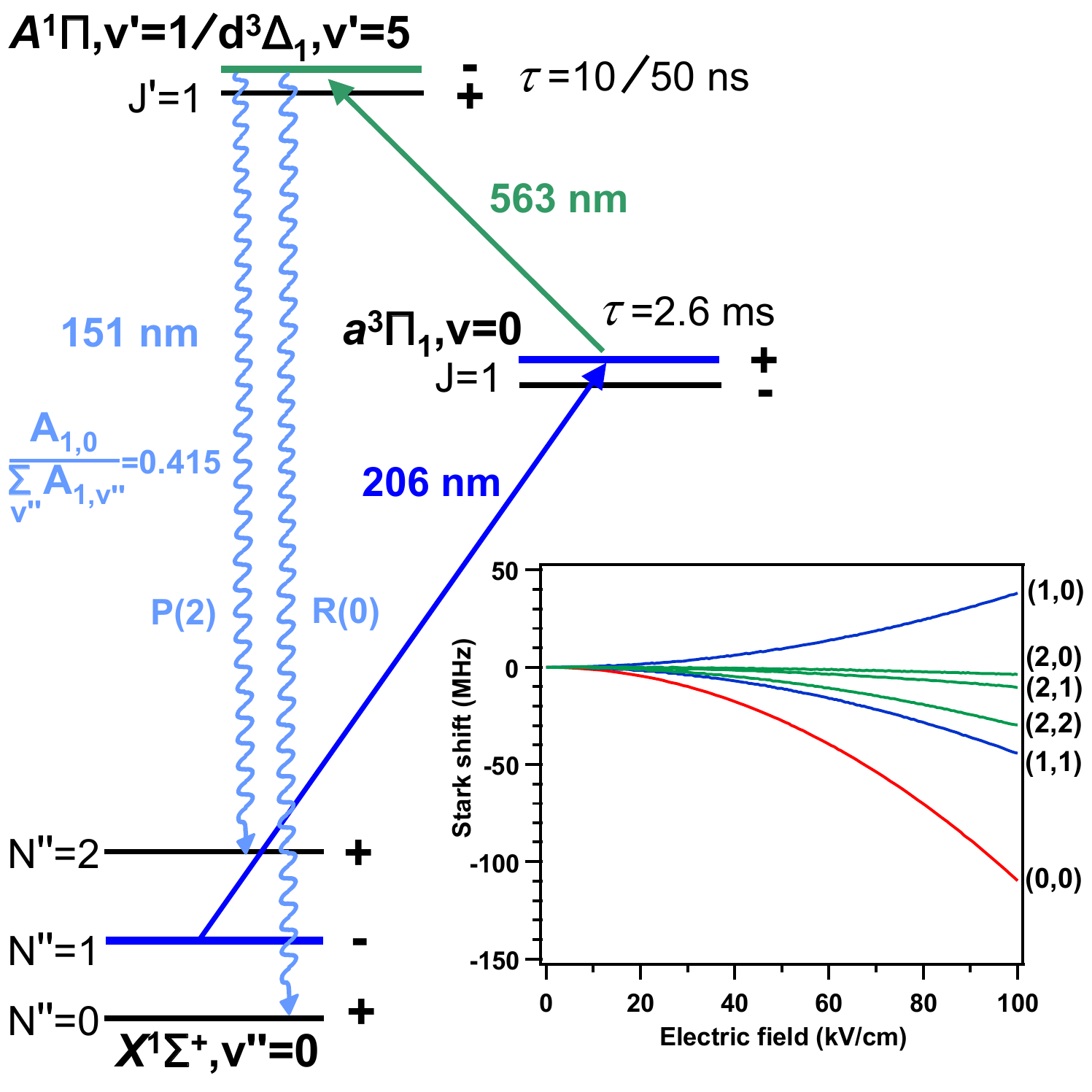}
 \caption{(color online) Relevant energy levels of $^{12}$C$^{16}$O involved in
 the unidirectional optical transfer demonstrated here.
 CO molecules are prepared in the a$^3\Pi_1$, $v$=0
 metastable state by laser excitation from the electronic
 ground state. The metastable CO molecules can be slowed down to low
 velocities using a Stark decelerator and can then be optically pumped to
 the $J'$=1 level in the short-lived A$^1\Pi$,~$v'$=1/d$^3\Delta_1$,~$v'$=5 state. Up
 to 28\% of the molecules in the metastable state can thus be transferred
 to the absolute ground level X$^1\Sigma^+$,~$v''$=0,~$N''$=0.
 The inset shows the Stark shifts for the lowest three rotational
 levels of ground-state $^{12}$C$^{16}$O (labelled as
 ($N''$,$|M''|$))}.
 \label{Energylevels}
\end{figure}

After the metastable CO molecules have been brought to a
standstill, they can be dumped back to the
X$^1\Sigma^+$,~$v''$=0,~$N''$=0 level via stimulated emission
pumping, thus producing the required translationally cold
ground-state molecules. Preferably, there should be dissipation in
the process of returning to the ground state, such that the
phase-space density can be increased by accumulating molecules
from subsequent cycles 
\cite{Stuhler-PRA-2001,Meerakker-PRA-2001,Falkenau-PRL-2011,Riedel-EPJD-2011}. 
Although the phosphorescence back to the electronic ground state could in
principle be exploited for this, its millisecond time scale makes
this impractical. It is shown here that it is possible to optically
pump the CO molecules from the a$^3\Pi$ state to perturbed levels in the A$^1\Pi$
state, and to use the fast spontaneous emission on the A$^1\Pi$
$\rightarrow$ X$^1\Sigma^+$ transition to populate the
ground-state level, as schematically shown in
Fig.~\ref{Energylevels}.

%--------------------- Fig 1 --------------------------------

From the negative parity $\Lambda$-doublet component of the $J'$=1
level in the A$^1\Pi$ state, emission can only take place on R(0)
and P(2) transitions, populating the $N''$=0 and $N''$=2 levels in
various vibrational levels of the electronic ground state,
respectively. We note here that the molecules in the $N''$=2 levels will be more difficult to trap, because of their smaller dipole moment. In spite of the five-fold degeneracy of the $N''$=2
levels, the R(0) transitions are twice as strong as the P(2)
transitions \cite{Hansson-JMS-2005}. The total fraction of the
spontaneous emission from the A$^1\Pi$, $v'$=1 level that ends up in
the X$^1\Sigma^+$, $v''$=0 level, given by the ratio
{$\sl A$$_{1,0}$}/$\sum_\mathrm{v''}${$\sl A$$_{1,\mathrm{v''}}$} of
Einstein-$\sl A$ coefficients, is 0.415.
This value is computed using the Rydberg-Klein-Rees (RKR) potentials
and the high-level \emph{ab initio} A $\leftarrow$ X transition
dipole function described elsewhere \cite{Gilijamse-JCP-2007}.
Therefore, if the excitation laser that drives the
A$^1\Pi$,~$v'$=1,~$J'$=1$\leftarrow$ a$^3\Pi_1$~$v$=0,~$J$=1
transition is intense enough and is on for a time that is long
compared to the approximately 10~ns radiative lifetime of the A$^1\Pi$ state
\cite{Field-JCP-1983}, up to 28\% of all the CO molecules in the
$J$=1 level of the a$^3\Pi_1$ state can be transferred to the single
$v''$=0, $N''$=0 level in the electronic ground state in this way. This transfer efficiency only holds when the parity of all the levels involved in the scheme is well defined. In an electric field the presence of states with mixed parity will result in leakage to the $N'=1$ level, decreasing the number of molecules ending up in the $N''=0$ level. 

The success of this scheme depends on the strength of the
spin-forbidden A$^1\Pi$,~$v'$=1 $\leftarrow$ a$^3\Pi$,~$v$=0
transition. With the RKR potentials and \emph{ab initio}
spin-orbit couplings from \cite{Gilijamse-JCP-2007}, a value of
$7.6 \times 10^{-4}$ D is computed for the intrinsic
transition dipole moment \cite{Gerrit}. However, the A$^1\Pi$
state has numerous perturbations, for instance resulting from interactions with the
$a'$$^3\Sigma^+$, d$^3\Delta$ and e$^3\Sigma^-$ states, which give
significant triplet character to selected rotational and
vibrational levels of the A$^1\Pi$ state. In $^{12}$C$^{16}$O, the
low rotational levels of the A$^1\Pi$, $v'$=1 state, in
particular, are perturbed by those of the d$^3\Delta_1$, $v'$=5
state, giving the A$^1\Pi$, $v'$=1, $J'$=1 level 17\% $^3\Delta_1$
character \cite{Morton-ApJ-1994}. Using a- and d-state RKR
potentials based on \cite{Gilijamse-JCP-2007} and
\cite{Field-JMS-1972} and a newly computed {\em ab initio} d-a transition
dipole function \cite{Gerrit}, a
transition dipole moment of 0.17 D is calculated for the
d$^3\Delta$, $v'$=5 $\leftarrow$ a$^3\Pi$, $v$=0 transition.
Therefore, the perturbation by the d$^3\Delta_1$ state makes the
A$^1\Pi$, $v'$=1, $J'$=1 $\leftarrow$ a$^3\Pi$, $v$=0, $J$=1
transition weakly allowed, and considerably stronger than 
expected based on its intrinsic transition dipole moment. 

To quantify the strength of this particular rotational transition further,
we have to consider the H\"{o}nl-London factors for the transitions 
from the a$^3\Pi_1$, $v$=0, $J$=1 level to the various rotational 
levels in the d$^3\Delta$, $v'$=5 state. Unfortunately, the transitions with
$\Delta \Omega$=0 are forbidden in first order approximation; transitions 
from the a$^3\Pi_1$ spin-orbit manifold are strongest when they end up 
in the d$^3\Delta_2$ spin-orbit manifold, but this does not couple to the
A$^1\Pi$ state. Using the available spectroscopic data for the a$^3\Pi$ 
state \cite{Field-JMS-1972} and the d$^3\Delta$ state \cite{Carroll-JCP-1962}, 
the transitions to the $J'=1$ and $J'=2$ levels in the d$^3\Delta_1$, $v'$=5 state 
are calculated to have a H\"{o}nl-London factor of only $1.0 \times 10^{-2}$ 
and $7.5 \times 10^{-2}$ (when the sum of all transitions from the $J$=1 
level is normalized to $2J+1=3$), respectively. Nevertheless, all together this still leads 
to an Einstein B-coefficient for the d$^3\Delta_1$, $v'$=5, $J'$=1 $\leftarrow$ 
a$^3\Pi$, $v$=0, $J=1$ transition that is, for instance, about an order of 
magnitude larger than that for the Q$_2$(1) transition in the Cameron band 
that is used to prepare the CO molecules in the a$^3\Pi_1$, $v$=0, $J$=1 
level. This implies that the optical pumping of CO molecules from the
metastable state to their absolute ground state via the $J'$=1 level in the 
mixed d$^3\Delta_1$, $v'$=5 // A$^1\Pi$, $v'$=1 system can be induced by
a cw laser operating around 563 nm.  

The existence of such "doorway" states in CO, via which transitions from
the singlet manifold to the triplet manifold can be efficiently induced, has
been known for a long time. In the early seventies already, emission from 
the mixed d$^3\Delta_1$, $v'$=5 // A$^1\Pi$, $v'$=1 system to excited
vibrational levels of the a$^3\Pi$ state was observed in the visible part of 
the spectrum after vacuum ultra-violet (VUV) excitation from the ground 
state \cite{Slanger-CPL-1970}. Perturbations of higher vibrational levels of 
the A$^1\Pi$ state with the a$'^3\Sigma^+$ state have also been used to populate levels in the 
metastable triplet state via stimulated emission pumping \cite{Sykora-JCP-1999}.
The A$^1\Pi$, $v'$=1, $J'$=1,2 $\leftarrow$ a$^3\Pi$, $v$=0, 
$J$=1 and d$^3\Delta_1$, $v'$=5, $J'$=1,2 $\leftarrow$ a$^3\Pi$, $v$=0, $J$=1
rotational transitions around 563 nm have been observed both via depletion of metastable CO 
signal \cite{Santambrogio-2010} and via recording of the VUV emission to 
the ground state. Using the latter method, we have determined the lifetime of the excited states and the Stark shift of the aforementioned transitions, from which the degree of parity mixing in any electric field can be calculated. 

\section{Experimental}

A pulsed beam of $^{12}$C$^{16}$O molecules with a mean velocity
of around 430 m/s is produced by expanding a mixture of 20\% CO in Xenon
from a valve at room temperature. The CO molecules are
excited to the upper $\Lambda$-doublet component of the a$^3\Pi_1$, 
$v$=0, $J$=1 level by direct laser excitation from the electronic ground 
state on the Q$_2$(1) transition (indicated by the blue arrow in 
Fig.~\ref{Energylevels}). The metastable CO molecules pass through
a Stark decelerator with 108 electric field stages, described in more detail 
elsewhere~\cite{Meerakker-ARPC}. In the present study, the decelerator is operated in the $s=3$ overtone guiding mode, producing a state-selected beam of metastable CO molecules with a mean velocity of 430 m/s. After exiting the decelerator the CO 
molecules are excited from the a$^3\Pi_1$ state to the lowest rotational 
levels in the d$^3\Delta_1$, $v'$=5 and A$^1\Pi$, $v'$=1 states, using
either a pulsed or cw dye laser system, crossing the molecular beam under
right angles. The pulsed laser (5 ns pulse duration) is used to record an
overview VUV spectrum and to determine 
the fluorescence decay times while the narrow-band cw ring dye laser 
system enables a high resolution measurement of the spectral line
profiles, and in particular of the Stark-splitting of the rotational transitions 
when an external electric field is applied. The cw ring dye laser is stabilized 
to a frequency-stabilized HeNe laser, resulting in a overall frequency 
resolution of 1~MHz, which is over an order of magnitude smaller than 
the spectral width of the peaks caused by residual Doppler and lifetime
broadening. The transmission fringes of an etalon (free spectral 
range of about 150 MHz) are measured simultaneously to determine the 
relative frequency. The VUV emission from the d$^3\Delta_1$, $v'$=5 and 
the A$^1\Pi$, $v'$=1 states to the ground state is measured using a solar 
blind photomultiplier tube (Electron Tubes type 9424B), placed further 
downstream on the molecular beam axis. A homogeneous electric field, with 
a direction perpendicular to both the molecular beam axis and the laser 
beam propagation direction, can be applied in the detection region. For this,
a voltage difference is applied to two circular electrodes (40~mm diameter) that 
are placed 6.9~mm apart. 

%--------------------- Fig 2 --------------------------------
\begin{figure}[h]
 \centering
     \includegraphics[width=7.5cm]{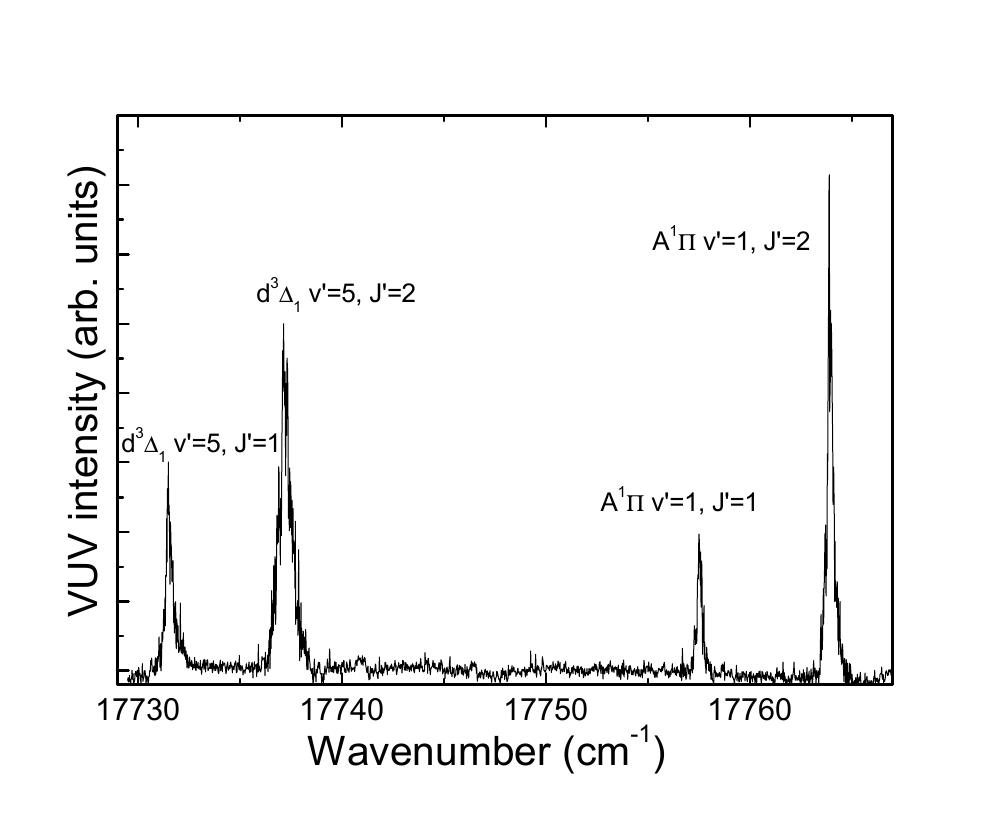}
 \caption{VUV excitation spectrum of metastable $^{12}$C$^{16}$O 
 molecules. Using a pulsed laser, the molecules are optically excited from the 
 upper (positive parity) $\Lambda$-doublet component of the a$^3\Pi_1$, 
 $v$=0, $J$=1 level to the levels indicated in the Figure, while the intensity 
 of the VUV fluorescence back to the ground-state is recorded.} 
 \label{OverviewSpectrum}
\end{figure}

An overview of the VUV excitation spectrum is shown in Fig.~\ref{OverviewSpectrum}. 
The four lines observed in the spectrum end up in the lowest two rotational levels 
of the mutually interacting d$^3\Delta_1$, $v'$=5 // A$^1\Pi$, $v'$=1 states. The 
transitions are saturated and power-broadened, and the excitation spectrum shown 
here merely serves the purpose of identifying the various spectral lines. As discussed
before, the intrinsic strengths of the transitions to the $J'$=1,2 rotational levels in the 
d$^3\Delta_1$, $v'$=5 state are about a factor of five stronger than the corresponding 
transitions to the A$^1\Pi$, $v'$=1 state and transitions to the $J'$=2 levels 
are about a factor 7.5 stronger than those to the $J'$=1 levels. In order of increasing 
frequency, the four transitions from the a$^3\Pi_1$, $v$=0, $J$=1 level thus have
relative line strengths of 1.0:7.5:0.2:1.5.

%--------------------- Fig 3 --------------------------------
\begin{figure}[h]
 \centering
     \includegraphics[width=7.5cm]{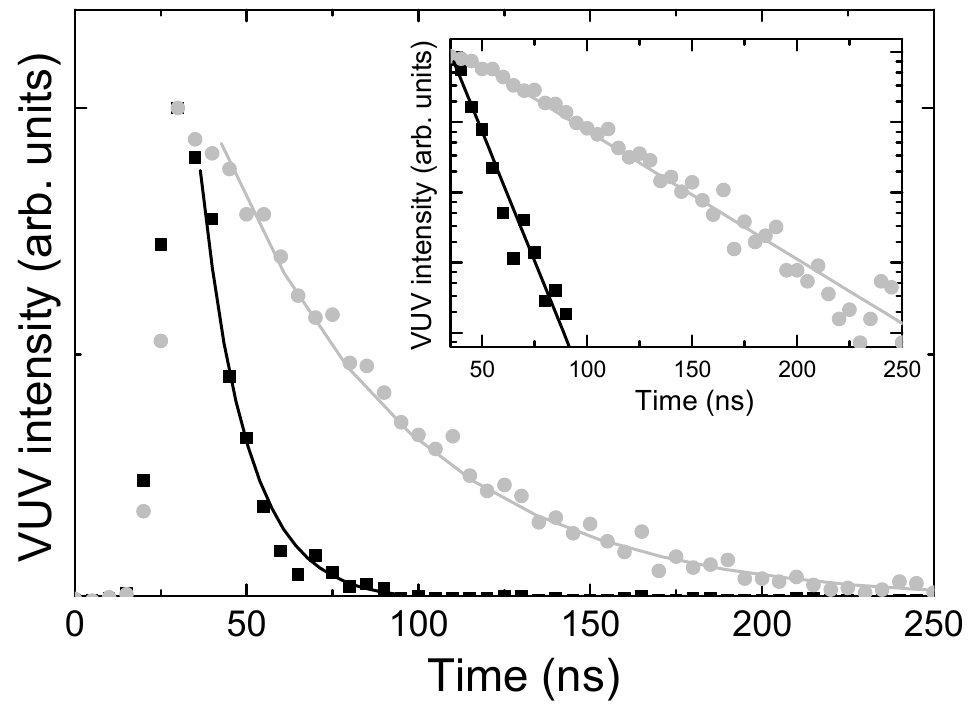}
 \caption{VUV fluorescence intensity as a function of time for CO molecules laser excited 
 to the negative parity component of the A$^1\Pi$, $v'$=1, $J'$=1 level (squares) and the 
 d$^3\Delta_1$, $v'$=5, $J'$=1 (dots) level. The solid curves result from a least-squares fit 
 to singly exponentially decaying transients. In the inset, the tails of the VUV fluorescence 
 data are shown on a logarithmic scale, together with the best fitting straight lines.}
 \label{Lifetimes}
\end{figure}

In Fig.~\ref{Lifetimes} the VUV fluorescence intensity is shown as a function of time
during and after pulsed 563~nm laser excitation to the $J'$=1 levels
of the A$^1\Pi$, $v'$=1 state (squares) and the d$^3\Delta_1$, $ v'$=5 state (dots).
The radiative lifetimes are extracted from the exponentially decaying tail of the VUV 
intensity after the laser excitation pulse is over. A weighed least-squares 
fit to a singly exponentially decaying curve in the time interval from 35 to 90~ns gives 
a value of $13.5 \pm 2.0$~ns for the A$^1\Pi$, $v'$=1, $J'$=1 level. As the data 
acquisition system only has a time resolution of 5~ns and as the laser intensity will
not drop off abruptly at the end of the pulse, this value is an upper limit for the
radiative lifetime of the A$^1\Pi$, $v'$=1, $J'$=1 level.
A similar fit in the time interval from 35 to 325~ns gives a radiative lifetime of 
$55 \pm 4$~ns for the d$^3\Delta_1$, $v'$=1, $J'$=1 level. As the radiative lifetime of 
unperturbed vibrational levels in the d$^3\Delta_1$ state is on the order of several 
micro-seconds \cite{Paske-JCP-1980} and thus much longer than that of the unperturbed 
A$^1\Pi$, $v'$=1 state, the latter, $\tau_A$, can be determined as the product of the 
two independently measured lifetimes divided by their sum, resulting in a value of 
$\tau_A$=$10.8 \pm 1.8$~ns. Within the error bar, this value agrees with the deperturbed 
value of the radiative lifetime of the A$^1\Pi$ $v'$=1 state as determined in the classic 
paper of Field {\it et al.} \cite{Field-JCP-1983}. The ratio of the two independently measured 
lifetimes is found as $4.1 \pm 0.7$, and is a direct measure for the fraction of A$^1\Pi$ 
state over $^3\Delta_1$ state character in the wavefunction of the A$^1\Pi$, $v'$=1 state. 
The latter is actually known with high precision from spectroscopic data 
as 4.88 \cite{Morton-ApJ-1994}, and these independent lifetime measurements 
are thus seen to be consistent with this. 

%--------------------- Fig 4 --------------------------------
\begin{figure}[h]
 \centering
     \includegraphics[width=7.5cm]{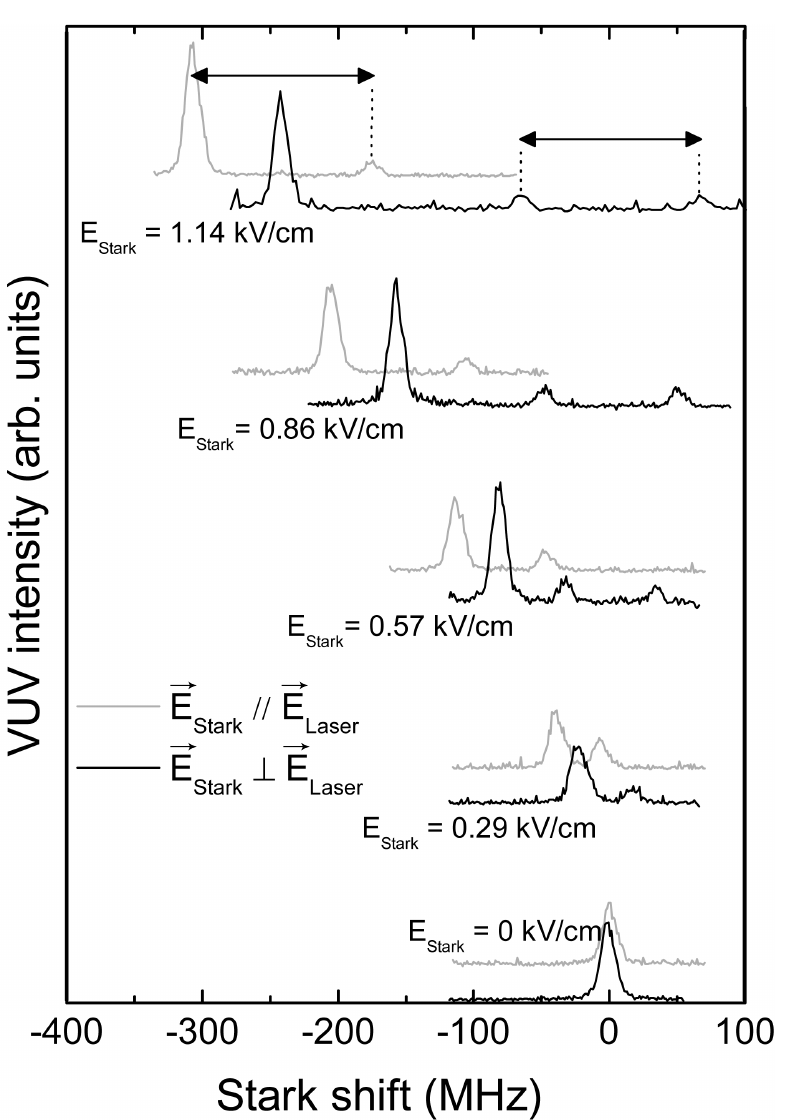}
 \caption{Spectrum of the d$^3\Delta_1$, $v'$=5,
$J'$=1 $\leftarrow$ a$^3\Pi_1$, $v$=0, $J$=1 transition
measured in electric fields up to 1.14~kV/cm. The black (grey)
curves correspond to the situation where the light is polarized
perpendicular (parallel) to the electric field.}
 \label{D3DeltaStark}
\end{figure}

To determine the electric dipole moment of the d$^3\Delta_1$, $v'$=5 state, we
have recorded the shifts and splittings of the transition to the $J'$=1 level in the
presence of an external electric field. The high-resolution Stark-shift spectra of 
the d$^3\Delta_1$, $v'$=5, $J'$=1 $\leftarrow$ a$^3\Pi_1$, $v$=0, $J$=1 
transition are shown in Fig.~\ref{D3DeltaStark} in electric fields up to 1.14~kV/cm 
for the excitation light polarized perpendicular (black curves) or parallel (grey
curves) to the applied electric field. The Stark shifts are given relative to the 
frequency of the field-free transition at 17731.7~cm$^{-1}$~\cite{Morton-ApJ-1994,Field-JMS-1972}.
In the metastable a$^3\Pi_1$, $v$=0 state only the upper $\Lambda$-doublet
component of the $J$=1 level is populated. Although only molecules in the low-field 
seeking $M \Omega$ = $-$1 component are actively guided through the decelerator,
a minor fraction of metastable CO molecules might be present in the $M$=0 level
in the interaction region as well. In $^{12}$C$^{16}$O these levels are degenerate
in zero electric field, and transitions from the low-field seeking component to the
$M$=0 level are difficult to avoid in regions of low electric 
field \cite{Meek-Science-2009}, \emph{e.g.}, near the exit of the decelerator. 

When the polarization of the laser ($\vec{E}_{Laser}$) is perpendicular to the direction of 
the electric field ($\vec{E}_{Stark}$), the selection rule for the optical transitions is given 
by $\Delta M$ = $\pm$1. In this case, we expect a single strong transition originating 
from the low-field seeking $M \Omega$ = $-$1 component in the metastable a$^3\Pi$ 
state to the $M$=0 level(s) in the d$^3\Delta_1$ state. In the d$^3\Delta_1$ state, the 
$\Lambda$-doubling is expected to be negligibly small, and the two $M$=0 levels 
(one of either parity) will thus be (nearly) degenerate. Furthermore, as these $M$=0 
levels will not shift at all in the relatively low electric fields that we have used in our 
experiments, the downward Stark shift of the strong transition as observed in 
Fig.~\ref{D3DeltaStark} can be solely attributed to the upward Stark shift of the $M \Omega$ = $-$1
component of the a$^3\Pi$, $v$=0, $J$ = 1 level. The latter is very well known, and the observed 
shift of this transition in the spectrum has therefore actually been used to accurately calibrate 
the applied electric fields, yielding the values given in the Figure. The two additional 
weak peaks that are observed in the spectrum for $\vec{E}_{Laser} \perp \vec{E}_{Stark}$
are the transitions from the $M$=0 level in the a$^3\Pi$ state to the $M \Omega$= $\pm$ 1 
levels in the d$^3\Delta$ state. The splitting between these peaks, indicated by
the solid arrow in the spectrum recorded in an electric field of 1.14 kV/cm, is
given by $(\mu_{\Delta} E_{Stark})/h$, where $\mu_{\Delta}$ is the magnitude of
the electric dipole moment in the d$^3\Delta_1$, $v'$=5 state, $E_{Stark}$ is the 
magnitude of the applied electric field and $h$ is Planck's constant. 

The same splitting is also observed, and indicated as such, when the polarization of the 
laser is parallel to the electric field. In this case, the selection rule for the optical transitions is
$\Delta M$ = 0; transitions between $M$=0 levels are forbidden. The two peaks that
are observed in the spectrum both originate from the low-field seeking 
$M \Omega$ = $-$1 component in the metastable a$^3\Pi$ state and go the 
$M \Omega$= $\pm$ 1 levels in the d$^3\Delta$ state. The magnitude of the observed
splitting only gives information on the magnitude of the dipole moment, but from the relative 
intensity of the two peaks information on the sign of the dipole moment can be extracted.
As mentioned before, the transition from the a$^3\Pi$ state to the d$^3\Delta_1$ state is
forbidden in first order approximation. It borrows transition dipole moment from the 
$\Delta \Omega$ = $+$1 transitions via the weak intramolecular interaction between rotational 
levels with different quantum number $\Omega$. In the presence of an electric field, it
borrows additional transition dipole moment due
to the Stark interaction between different $J$ levels in both the a$^3\Pi$ and d$^3\Delta$ 
state; the observed electric field dependence of the (relative) intensity of the two peaks 
actually indicates that the latter effect is quite strong. 
The interference between the various contributions to the transition dipole moment can 
be constructive or destructive, depending on the mutual sign of the permanent electric 
dipole moment of the a$^3\Pi$ and d$^3\Delta$ state. The observed intensity ratio of the 
two peaks, and the way in which this ratio changes with increasing electric field, can only 
be quantitatively explained if the electric dipole moment in the d$^3\Delta$, $v'$=5 state 
has a different sign than the dipole moment in the a$^3\Pi$ state. From the complete 
analysis of the spectra shown in Fig.~\ref{D3DeltaStark} we thus deduce a value for the 
electric dipole moment in the $J''$=1 level of the d$^3\Delta_1$, $v'$=5 state of 
$-$(0.23$\pm$0.01) D. 

The electric dipole moment of CO in the $v'$=3, $v'$=4 and $v'$=6 levels of the d$^3\Delta$ 
state has been determined earlier as $-$0.48 D, $-$0.42 D, and $-$0.28 D, respectively
\cite{Lisy-JCP-1979}. From a linear interpolation of these values, one would have expected
a value of around $-$0.35 D for the dipole moment in the $v'$=5 level. Due to the perturbation
with the A$^1\Pi$, $v'$=1 state, however, this value is slightly reduced, as will be discussed in 
more detail later.

%--------------------- Fig 5 --------------------------------
\begin{figure}[h]
 \centering
     \includegraphics[width=7.5cm]{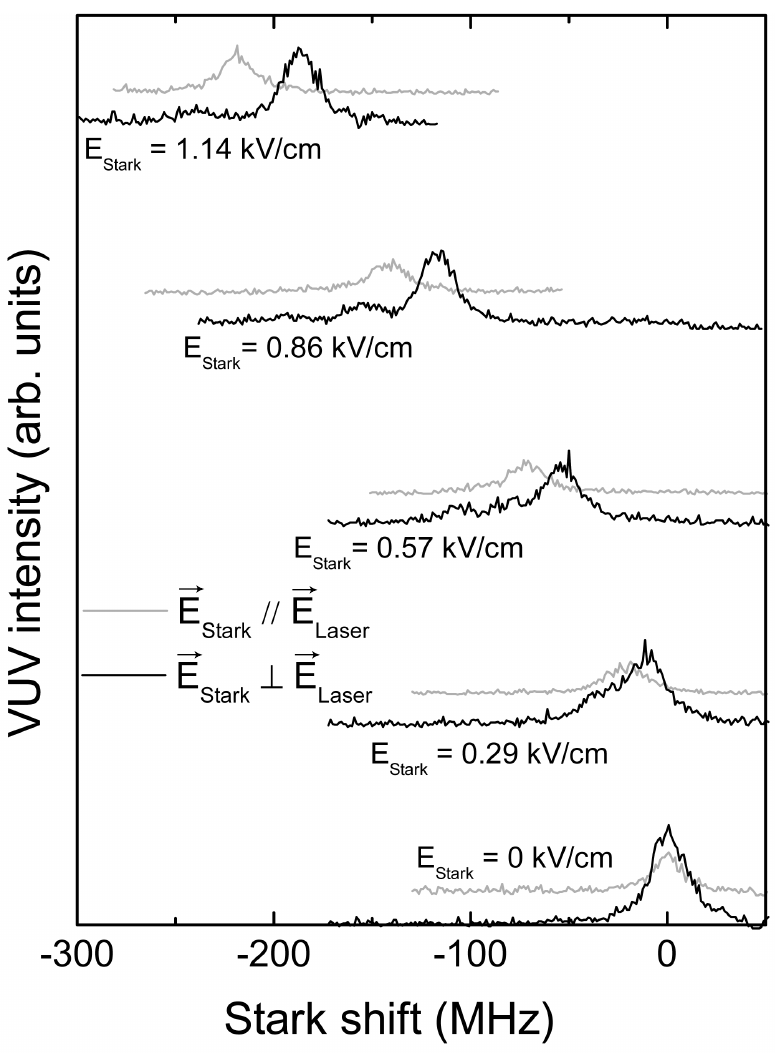}
 \caption{Spectrum of the A$^1\Pi$, $v'$=1, $J'$=2 $\leftarrow$ a$^3\Pi_1$, $v$=0, $J$=1 
 transition measured in electric fields up to 1.14~kV/cm. The black (grey) curves correspond 
 to the situation where the light is polarized perpendicular (parallel) to the electric field.}
 \label{A1PiStark}
\end{figure}

To determine the electric dipole moment of the A$^1\Pi$, $v'$=1 state, we
have recorded the shifts and splittings of the transition to the $J'$=2 level in the
presence of an external electric field; the transition to the $J'$=1 level that we would 
have preferred to use is unfortunately too weak for this. The high-resolution 
Stark-shift spectra of the A$^1\Pi$, $v'$=1, $J'$=2 $\leftarrow$ a$^3\Pi_1$, $v$=0, $J$=1 
transition are shown in Fig.~\ref{A1PiStark} in electric fields up to 1.14~kV/cm 
for the excitation light polarized perpendicular (black curves) or parallel (grey
curves) to the applied electric field. The Stark shifts are given relative to the 
frequency of the field-free transition at 17763.9~cm$^{-1}$~\cite{Morton-ApJ-1994,Field-JMS-1972}.
In these experiments, only the low-field seeking $M \Omega$ = $-$1 component of 
the $J$=1 level in the a$^3\Pi_1$, $v$=0 state appeared to be significantly populated.

For the A$^1\Pi$ state, it is not expected that the $\Lambda$-doubling can be neglected
and the magnitudes of both the $\Lambda$-doubling and the dipole moment need to be
extracted from the observed Stark shifts. The observed shifting of the transitions to lower
frequency with increasing electric field is again mainly due to the increase in Stark
energy of the level in the a$^3\Pi$ state. The least shifted component is
observed when $\vec{E}_{Laser} \perp \vec{E}_{Stark}$, \emph{i.e.}, when $\Delta M$ = $\pm$1 such
that the most upward shifting $M \Omega$ = $-$2 component in the A$^1\Pi$ state can be 
reached. The shoulder
at slightly larger red-shift is due to the transition to the $M$=0 component. When the
polarization of the laser is parallel to the direction of the electric field, the $M \Omega$ = $-$1
component can be reached, and this transition is indeed seen to be spectrally located in 
between the two transitions just mentioned. It turns out that in this case the observed intensities 
can not only be used to determine the sign of the dipole moment, but also to unambiguously
establish that the upper (lower) $\Lambda$-doublet component of the $J'$=2 level in the
A$^1\Pi$, $v'$=1 state has negative (positive) parity. Only in this case the observed intensity
pattern is quantitatively reproduced and, in particular, the factor two intensity difference in the 
strength of the transition to the $M \Omega$ = $-$2 component compared to the transition to 
the $M \Omega$ = $-$1 component is obtained. From a fit of the observed Stark shifted patterns, 
a value of the $\Lambda$-doublet splitting between the parity levels under field-free conditions 
of 18 $\pm$ 3 MHz is obtained. The electric dipole moment in the $J'$=2 level of the 
A$^1\Pi$, $v'$=1 state is determined to be $+$(0.29$\pm$0.02) D.

\section{Conclusions}

In this paper, we have outlined and demonstrated an experimental scheme to
produce samples of translationally cold CO molecules in their absolute ground-state level. 
The scheme is based on the use of the remarkably efficient unidirectional optical transfer 
from the metastable a$^3\Pi$ state to the $N'$=0 level in the X$^1\Sigma^+$, $v''$=0 state 
\emph{via} the $J'$=1 level in the perturbed d$^3\Delta_1$, $v'$=5 // A$^1\Pi$, $v'$=1 system.
The radiative lifetime of the $J'$=1 level in the (nominally) d$^3\Delta_1$, $v'$=5 state is indeed
seen to be as short as 55 $\pm$ 4 ns due to the coupling with the singlet state. Fluorescence 
from this rotational level in the triplet manifold will end up in the singlet manifold, in particular
in the X$^1\Sigma^+$ electronic ground state, with a quantum efficiency that is very close to 
unity (about 0.99). The transition to this level from the metastable state can be efficiently 
induced by a cw laser; in the present experiments about 300~mW of narrowband 563 nm 
laser radiation in a two millimeter diameter beam has been used to drive this transition. 

As the manipulation of the translational motion of the metastable CO molecules relies on
their interaction with electric fields, it is important to know the dipole moments of the 
intermediate levels in the optical transfer scheme as well. The dipole moment of the $J'$=1
level in the d$^3\Delta_1$, $v'$=5 state is found to be $-$(0.23$\pm$0.01) D. The dipole 
moment in the perturbing A$^1\Pi$, $v'$=1 state has the opposite sign; for the $J'$=2 level 
it is found to be $+$(0.29$\pm$0.02) D. From a detailed analysis of the perturbation between 
these states, it is known that the amount of $^1\Pi$ character in the wavefunction of the
d$^3\Delta_1$, $v'$=5, $J'$=1 level is 17\% whereas in the wavefunction of the $J'$=2 
level in the A$^1\Pi$, $v'$=1 state it is 84.3\% \cite{Morton-ApJ-1994}. The deperturbed 
value of the dipole moment of the d$^3\Delta$, $v'$=5 state can then be calculated as 
$-$(0.36$\pm$0.02) D, whereas the deperturbed value for the A$^1\Pi$, $v'$=1 state
results in $+$(0.41$\pm$0.03) D. The first value falls nicely in between 
the values found for the adjacent (unperturbed) vibrational levels of the d$^3\Delta$ state
\cite{Lisy-JCP-1979}. The deperturbed value of the dipole moment for the $v'$=1 level
in the A$^1\Pi$ state is somewhat higher than the value of $+$(0.335 $\pm$ 0.013) D that
has been found for the $v'$=0 level \cite{Drabbels-JCP-1993}. A slight increase of the
permanent electric dipole moment with increasing vibrational quantum number is indeed
expected from the calculated dipole moment function for the A$^1\Pi$ state \cite{Cooper-JCP-1987}. \\

\section{Acknowledgments} 

We acknowledge useful discussions with S.Y.T. van de Meerakker, S.A. Meek,
and W. J\"{a}ger. J.H.B. thanks the Alexander von Humboldt Foundation for a 
post-doctoral fellowship. This work has been financially supported by the 
ERC-2009-AdG program under grant agreement 247142-MolChip.

\end{document}